# Is $Pr_{1-x}Ca_xMnO_3$ a ferroelectric of new mechanism?

Shuai DONG[1] and J.-M. LIU[2]

(22 Oct. 2006)

*Nanjing National Laboratory of Microstructures, Nanjing University, Nanjing 210093, China
International Center for Materials Physics, Chinese Academy of Sciences, Shenyang, China*

**In the Issue 12, Volume 3, a new type of charge-ordering pattern for $Pr_{1-x}Ca_xMnO_3$: a hybridization of site- and bond-centred ordering (SBCO)[1] was proposed by Efremov *et al.*. It was believed this discovery opened a new route to magnetic ferroelectrics[2]. Here we argue that the possible ferroelectric of $Pr_{1-x}Ca_xMnO_3$ also originates from the cation displacement, similar to prototypical ferroelectric system.**

Recently, the magnetic ferroelectrics attract more and more researchers' attention due to their special coupling between magnetism and ferroelectricity. Some charge-ordering transition metal oxides are preferred candidates because of their special mechanism of ferroelectricity which avoids the incompatibility with magnetism. For these materials, the dipoles caused by charge disproportionation of identical metal cations contribute the polarization. A typical case is $LuFe_2O_4$, which has a bilayer structure along *c*-axis[3,4]. Its low temperature electric polarization comes from the dipoles consisted of the upper $Fe^{2+}$ layer and the lower $Fe^{3+}$ layer.

It is known that the nonzero misalign of positive and negative charge centres accounts for ferroelectricity. According to this criterion, the SBCO with unshifted Mn cations contributes zero to net polarization. In fact, the estimate of the spontaneous polarization in ref. 1 was determined by the difference of the Mn-Mn distances within and between dimers, which implied that the possible polarization also came from the shift of Mn cations (i.e. distortion of lattice), which is similar to prototypical ferroelectric system. However, the shift of Mn against surrounding O octahedron is beyond the DDEX model in ref. 1.

As stated in in ref. 1, SBCO dimer breaks the inversion symmetry which may distort the lattice in real system. Even though, it is hard to observe the polarization in $Pr_{1-x}Ca_xMnO_3$. The dipole in charge-ordering materials is a dimer of two neighbouring cations. Therefore, the net polarization will disappear if the dimers on material surface are truncated. Contrarily, this problem does not exist in $BaTiO_3$ whose dipole is not dimer. And the bi-layered character of $LuFe_2O_4$ preserves the $Fe^{2+}$-$Fe^{3+}$ dipoles when the surface is the cross section between layers. It is why the spontaneous polarization was observed in $LuFe_2O_4$[3].

---

1  Email: saintdosnju@gmail.com
2  Email: liujm@nju.edu.cn

In conclusion, the possible polarization of $Pr_{1-x}Ca_xMnO_3$ also originates from the conventional cation shift in O octahedron. Therefore, it can not avoid the incompatibility between ferroelectricity and magnetism.